\documentclass[useAMS,usenatbib]{mn2e}

\usepackage[utf8]{inputenc}
\usepackage[english]{babel}
\usepackage{array}

\usepackage{natbib}
\usepackage{graphicx}
\usepackage{amsmath}
\newcommand{\fas}{\mbox{\ensuremath{.\!\!^{\prime\prime}}}}
\newcommand{\uJy}{\,\ensuremath{\umu}\text{Jy}{}}

\title[Stacking of large interferometric data sets in the image- and uv-domain -- a comparative study]{Stacking of large interferometric data sets in the image- and uv-domain -- a comparative study}
\author[L. Lindroos, K. K. Knudsen, W. Vlemmings, J. Conway, and I. Mart{\'i}-Vidal]{L. Lindroos$^{1}$\thanks{E-mail:
lindroos@chalmers.se}, K. K. Knudsen$^{1}$, W. Vlemmings$^{1}$, J. Conway$^{1}$, and I. Mart{\'i}-Vidal$^{1}$ \\
$^{1}$Department of Earth and Space Sciences, Chalmers University of Technology, Onsala Space Observatory, SE-439 92 Onsala, Sweden}
\begin{document}

\date{Accepted 2014 November 3. Received 2014 October 28; in original form 2014 January 20}

\pagerange{\pageref{firstpage}--\pageref{lastpage}} \pubyear{2014}

\maketitle

\label{firstpage}

\begin{abstract}
	We present a new algorithm for stacking radio interferometric data in the uv-domain.
	The performance of uv-stacking is compared to the stacking of fully imaged data
	using simulated Atacama Large Millimeter/sub-millimeter Array (ALMA) 
	and the Karl G. Jansky Very Large Array (VLA) deep extragalactic surveys.
	We find that image- and uv-stacking produce similar results,
	however, uv-stacking is typically the more robust method.
	An advantage of the uv-stacking algorithm is the availability of uv-data post stacking,
	which makes it possible to identify and remove problematic baselines.
	For deep VLA surveys uv-stacking yields a
	signal-to-noise ratio that is up to 20 per cent higher than image-stacking.
	Furthermore, we have investigated stacking of resolved
	sources with a simulated VLA data set where 1.5\arcsec (10--12\,kpc at $z\sim\,1-4$\,) sources are stacked.
	We find that uv-stacking, where a model is fitted directly to the visibilities,
	significantly improves the accuracy and robustness of the size estimates.
	While scientific motivation for this work is studying faint, high-$z$ galaxies,
	the algorithm analysed here would also be applicable in other fields of astronomy.
	Stacking of radio interferometric data is also expected to play a big role for future surveys with telescopes such as LOFAR and Square Kilometre Array (SKA).
\end{abstract}


\section{INTRODUCTION}
The last decades have seen a massive improvement in telescopes, especially for continuum sensitivity. 
With the advent of each new telescope, we push further and are able to study ever fainter sources.
But no matter the quality of the telescope, there will always be sources which are just beyond reach.
A method which can be used to extend the reach of a telescope further is so called \emph{stacking}.
Stacking relies on the fact that we already know the location of the sources from another observation.
It works through averaging the signal from several sources below the noise and allows statistical detection in the new data.
Stacking was originally developed as a method for optical instruments observing visible light \citep{cady1980}.
Later this has been expanded to be used at many different wavelengths,  \citep[e.g.][]{brandt01,bartelmann03,webb03,knudsen05,dole06}.
In this paper, we investigate stacking techniques for interferometric (sub-)mm/radio data
and discuss on the particulars required for stacking of these data.

Stacking techniques have proven to be a powerful method, 
as an example for deep multi wavelength high redshift galaxy surveys, such as
COSMOS \citep{scoville2007} and Extended Chandra Deep Field-South (ECDF-S, \citet{lehmer2007}).
These surveys are accompanied by deep Very Large Array (VLA) surveys, COSMOS \citep{schinnerer2007} and ECDF-S \citep{miller2008, miller2013}, 
reaching sensitivities around 10\uJy/beam at 1.4\,GHz.
Assuming galaxies are not resolved this corresponds to a luminosity $10^{23}$\,W\,Hz$^{-1}$ at $z=2$
\footnote{Using a standard cosmology with $H_0 = 67.3$\,km\,s$^{-1}$\,Mpc$^{-1}$, $\Omega_\Lambda = 0.685$ and $\Omega_m = 0.315$ \citep{planck2013}.}
(or star formation rate of $\sim200$\,M$_\odot{}$\,yr$^{-1}$, \citealt{condon}),
limiting us to study only the most extreme objects beyond redshift 2.
By using stacking, e.g. 3000 galaxies in the ECDF-S, the noise can be pushed down to around 0.2\uJy/beam \citep{carilli2008}.
This allows us to go significantly deeper;
in the case of \cite{carilli2008}, down to a star formation rate of $\sim$9\,M$_\odot$\,yr$^{-1}$ at $z=3$.
While the main scientific motivation for the present work is to study the radio and sub-millimetre emission of 
less extreme high redshift galaxies, the general methods are applicable in many areas of astronomy,
e.g. looking for faint radio-emission from large grains around young stars \citep{greaves2012}.

Unlike an optical telescope, an interferometric telescope does not image the sky directly.
Instead, it samples the Fourier transform of the brightness distribution of the source being observed.
For each pair of antennas, we measure one point for each integration, frequency and polarisation.
This measured value is a complex number and is called a \emph{visibility}.
The location of each visibility in the Fourier space is determined by the separation of the antennas.
This is usually denoted as $(u,v,w)$ where $(u,v)$ are the projected separation in the west and north direction, as seen from the source,
and $w$ is in the direction orthogonal to $(u,v)$.
The Fourier space is often referred to as the $uv$-plane and this term will be used in this paper.
With an array,
each pair of antennas will trace out an elliptical track in the $uv$-plane as Earth rotates.
The nature of this sampling leaves holes in the $uv$-plane.
To produce an image,
we are required to interpolate the visibilities in these holes 
and this may lead to artefacts in the final image.
For more details,
see e.g. \cite{thompson}.

Generally, stacking of interferometric data has been performed on the reconstructed images,
rather than directly on the measured visibilities. 
We have developed an algorithm to stack directly in the $uv$-plane.
In this paper, we compare the new algorithm to stacking in the image domain.
To do this, we use simulated data sets to test various aspects of stacking.
The data sets are simulated Atacama Large Millimeter/sub-millimeter Array (ALMA) 
and The Karl G. Jansky Very Large Array (VLA) observations.
We do this in the context of high redshift galaxy surveys.


The paper is structured as follows.
In \S \ref{sec:simulations}, we describe the methods used to generate the simulated data sets that are used to test the stacking algorithms.
The algorithms used to stack are described in \S \ref{sec:algorithms}, 
specifically stacking in the image domain in \S \ref{sec:imagestack},
and stacking in the uv domain in \S \ref{sec:uvstack}.
In \S \ref{sec:results} we present the results of stacking our simulated data sets.
Finally, in \S \ref{sec:discussion}, we discuss the implication for stacking.

\section{SIMULATED DATA}
\label{sec:simulations}

Simulated data provides us with a powerful tool to test the performance of our stacking algorithms
and to compare their prospective advantages.
We choose to make simulations of VLA and ALMA type of data,
this allows us to test stacking at the few GHz and sub-mm regime.
For this purpose we generate 11 different data set types simulating observations with these telescopes
with varying setup.
In this section, we discuss these simulations,
and how they are generated.
A brief description of each data set type is listed in table \ref{tab:sim}.
Simulation type 1-7 deal with VLA simulations, 
primarily with dynamic range issues, 
which are significant for deep VLA surveys.
Simulation type 8-10 deal with ALMA simulations,
where the smaller field of view reduces the importance of dynamic range,
but introduces the requirements of using multiple pointings.
These simulations deal primarily with the issues of stacking data
in mosaic maps.


The simulated data are generated using CASA version 4.2 with code based on the task \emph{simobserve}. 
It takes input of a model sky, a list of sources distributed across the field of view of the observation. 
From this it produces a measurement set,
the standard visibility data format for CASA.

\subsection{Model sky}
The input sky model consists of a number of randomised sources.
For each data set type sources are added from two different populations with their respective distributions.
A \emph{bright source population} that ensures that the distribution of all the sources resemble that of the real sky.
And a \emph{target population}, that is, the sources intended to be stacked.
The presence of bright sources in the data is important as they contribute to the noise.
They will typically be associated with side lobes that can not fully be removed.

The sources are introduced directly into the visibilities 
\begin{equation}
	\label{eq:uvmod}
	V_\mathrm{model} = \sum_k \frac{A_N(l_k, m_k) S_k}{n_k}  e^{-\left(\pi r_k \frac{u+v}{\lambda} \right)^2}
	e^{ 2\pi{i \frac{u l_k + v m_k + w(n_k-1)}{\lambda }} }
\end{equation}
where $(l_k,m_k)$ are the direction cosines relative to the phase centre,
$(u,v,w)$ are the projected separations in m,
$S_k$ is the flux density,
and $r_k$ is the size in radians,
all for source $k$.
The factor $A_N (l_k, m_k)$ is the unit-less primary-beam attenuation in direction $(l_k, m_k)$.
To calculate the primary-beam attenuation we use the built-in primary-beam models in CASA 4.2. 
Point sources are introduced by setting $r_k = 0$.
Extended source are limited by this definition to Gaussian,
no more complicated morphologies for individual sources are simulated.

The coordinates of the sources are randomized uniformly within a field size set for each data set type.
All fields are centred at 3h49\arcmin10\fas99, $-30^\circ00\arcmin00\fas00$ in J2000.
The declination $-30^\circ$ was chosen to be similar to the declination of the Extended Chandra Deep Field-South (ECDFS). 
The right ascension was chosen at random as it does not impact on the data.

\subsection{VLA}
Data set type 1-7 simulates various aspects of stacking in VLA observations.
The data set types are made increasingly more complex to simulate more realistic data sets.
Through this we can test the effects caused by different aspects individually.
All the simulations are carried out with VLA in A configuration, 
with 27 dishes and baselines up to 36\,km.
The observations are generated at a frequency of 1.4\,GHz with a bandwidth of 250\,MHz,
split into 10 channels.
Each visibility has an integration time of 10 s with each data set having a total integration time of 9.4 h.
To each visibility we add a random flux density with a Gaussian distribution with $\sigma = 10\,$mJy to both the real and imaginary part.
Accounting for integration time, bandwidth and number of baselines, 
this corresponds to a thermal noise limit of 2.1\,\uJy{}/beam in the centre of the pointing.

\subsubsection{Basic case}
Data set type 1 replicates a basic VLA observation.
All sources in the model are located within the inner $5\arcmin\times5\arcmin{}$ area of the VLA field of view.
This ensures that smearing is minimal and wide-field effects are less severe.
The model sources consist of 1 source at 10\,mJy and 25 sources at 6\,\uJy{}.
All sources are introduced as point sources, i.e.,
$r_k = 0$.

The data are imaged using Multi-Frequency Synthesis (MFS, \citealt{conway90}) combined with w-projection \citep{cornwell2008}.
The 10\,mJy source is deconvolved using CLEAN \citep{Hogbom1974} down to a threshold of 20\,\uJy. 
The image is produced with a pixel size of 0\fas25 (13$\times$6 pixels over the $3\fas2\times1\fas6$ beam) and an image size of 1440$\times$1440 pixels.
The w-projection used 256 planes and the final image is primary beam corrected.
Using the CLEAN model of the bright source a residual $uv$-data set is produced by subtracting the model from each visibility. 

The final primary beam corrected and deconvolved image achieves a centre noise of 2.3\,\uJy/beam.

\subsubsection{Sub-pixel sampling}
Due to numerical limitation, 
large surveys in the radio are often imaged with fewer pixels across the beam than is the case for data set 1.
Data set type 2 aims to test the effect of this on stacking.
The data sets of this type are generated identically to data set type 1,
except for in imaging a pixel size of 0\fas5 (3 pixels across the minor axis of the beam) is used.
The image size is also decreased to 720$\times$720 pixels to cover the same field.

\subsubsection{Wide-field effects}
Data set type 3 is based on data set type 2, 
but increases the area the model sources are spread over to $30\arcmin\times30\arcmin$,
close to the full field of view of VLA at 1.4\,GHz.
The number of 10\,mJy sources is increased from 1 to 10,
while the number of target sources is increased from 25 to 900.
To achieve a similar signal-to-noise after stacking the flux densities of the target sources are decreased to 1.0\,\uJy.

Also the imaging is changed to cover a larger field.
The pixel size is kept at 0\fas5,
but the image size is increased from 720$\times$720 to 4000$\times$4000 pixels.

\subsubsection{Varying the flux density distribution of the target sources}
Data set type 4 is based on data set type 3,
however, it changes the flux density distribution of the target sources.
The new flux density of data set type 4 is inspired by flux density distribution of Lyman Break Galaxies (LBG).
The flux densities of the target sources are randomized using a differential source density ($dN/dS$) derived from a Schechter function:
\begin{equation}
	dN/dS = \left( \frac{n_*}{S_*} \right) \left( \frac{S}{S_*}\right)^{-\alpha} e^{-\frac{S}{S_*}}
	\label{eq:schechter}
\end{equation}
where $\alpha = -1.6$, $S_* = 1.7$\uJy{}, $S$ is the source flux density, and $n_*$ is scaled such that the total number of target sources in each the data set becomes 900.
For $S$ outside the interval of 0.25$\,S_*$ and 10$\,S_*$ the differential source density is set to 0.
This is guided by the fact that the ultraviolet  emission LBGs at $z=3$ follow a 
Schechter function as shown by \cite{steidel1999}.
The relation by \cite{steidel1999} is rescaled assuming that both UV and radio continuum traces star formation leading to Equation \ref{eq:schechter}.
This results in an average flux density for the target sources of 0.99\,\uJy,
very close to the flux density for data set type 3.

The deconvolution threshold is increased to 50\,\uJy{} from 20\,\uJy{}.
The deconvolved image achieves a centre noise of 2.7\,\uJy{}/beam in the centre of the map.
This results in that the brightest target sources are around 4$\sigma$.
These are not deconvolved (deconvolution threshold of 50\,$\uJy/\mathrm{beam}${} compared to peak brightness of 10.8\,\uJy{}/beam for the brightest target sources)
and are still included in the coordinate list for stacking.

\subsubsection{Varying the flux density distribution of the bright sources}
In data set type 1-4 the deconvolution of the bright sources is simplified by the fact that all bright sources have the same flux density.
Data set type 5 aims to produce a more realistic data set with a more complicated flux density distribution.
However, all sources are still introduced as point sources.
The data set type is based on data set type 4,
but the bright sources are generated using the flux density distribution from \cite{bondi2008} derived from the COSMOS field.
This distribution is a polynomial in log-log space, i.e.
\begin{equation}
	\log\left[  \frac{(dN/dS) / (S^{-2.5})}{\mathrm{Jy}^{1.5}\,\mathrm{sr}^{-1}} \right] = \sum_{n=0}^6 a_n \left[\log(S/\mathrm{mJy})\right]^n
\end{equation}
where $S$ if the source flux density, $dN/dS$ is differential source density (in sr$^{-1}$\,Jy$^{-1}$), 
$a_0 = 0.805$, $a_1 = 0.493$, $a_2 = 0.564$, $a_3 = -0.129$, $a_4 = -0.195$, $a_5 = 0.110$, and $a_6 = -0.017$.
The number of bright sources is fixed to 417 for the data sets of type 5.
Only bright sources above 10\,\uJy{} are deconvolved.

Otherwise imaging and production of residual measurement set is done in the same manner as for data set type 3.
This results in a noise in the centre of the deconvolved image of 2.5\,\uJy/beam,
approximately 75 per cent above the thermal noise limit.

\subsubsection{Extended bright sources}
In data set type 6 the bright sources are given an angular extent.
This complicates the deconvolution of the bright sources and results in a higher, not fully Gaussian noise in the residual data.
The source size $r_k$ is randomised uniformly between 0\fas5 and 5\fas0 for each bright source.
Imaging, deconvolution, and production of residual measurement set is done in exactly the same way as for data set type 5.
This results in a noise in the centre of the deconvolved image of 2.9\,\uJy/beam.

\subsubsection{Extended target sources}
Data set type 7, the final data set type for VLA, aim to test the effect of stacking marginally extended target sources.
It is based on data set type 6, but all target sources are given a size $r_k = 1\fas5$ and a flux density of 2.96\uJy.

\subsection{ALMA}
Data set type 8-10 simulates various aspects of stacking in ALMA observations.
We use an ALMA Cycle 1 configuration, with 32 dishes and a maximum baseline of 300\,m,
similar to the maximum baselines in the ALESS survey \citep{hodge2013},
resulting in a resolution of approximately 1\fas2.
The centre frequency of all simulated observations are 230\,GHz and the bandwidth is 4\,GHz,
split over 100 channels.
The ALMA observation are all mosaics with multiple pointings.
Each visibility has an integration time of 6\,s and each pointing
has a total integration of 360\,s.
Noise is calculated using the noise models in CASA for the ALMA site, 
with a perceptible water vapour of $2\,$mm.

Compared to VLA the field of view for ALMA is small due to the higher frequency.
All our simulated ALMA observations use multiple pointings to increase the
field of our observations.

\subsubsection{Contiguous mosaic}
Data set type 8 consists of a contiguous ALMA mosaic spaced at a distance of 20\arcsec.
Using 22 pointings it covers an area of $100\arcsec\times100\arcsec$.
The model sources consist of 10 sources at 30\,mJy and 50 sources at 1\,mJy.
The fainter 50 sources serve as target sources.

The ratio between the flux density of the brightest sources and the target sources is 30.
For our VLA data set type this ratio is typically around 5000.
As such bright sources will not contribute as much to the noise in data set type 8
as compared to data set type 1-7.
Also note that a data set of type 8 contains 10 bright sources spread over 22 pointings.
This means that some pointings will contain target sources but no bright sources,
further decrease the noise contribution from bright sources.

To image data set type 8 we use MFS with mosaic gridding.
This allows to image all pointings in parallel and handle overlap.
We use a pixel size of 0\fas2 and an image size of 800$\times$800 pixels.
All sources above 2\,mJy are deconvolved down to a threshold of 2\,mJy using CLEAN.
The image is then primary beam corrected.
Using the CLEAN model of the bright sources a residual $uv$-data set was produced by subtracting
the model from each visibility. 
The final image has a noise of 0.4\,mJy/beam in the centre.

\subsubsection{Non-contiguous mosaic}
Data set type 9 is designed to be similar to ALMA surveys such as \cite{hodge2013}.
This differs from data set type 8 in the fact that the full field is not covered with pointings.
Instead, all pointings are centred on the bright sources in the field.
The target sources are scattered around the bright sources,
no further than $11\fas2$ away from the centre of the pointing.
This design is configured to emulate non-contiguous mosaics centred on known sources,
which would typically be $50$ times brighter than the target stacking sources.
Note here that this still results in a small dynamic range compared data set type 1-7.
We again use a model with 10 sources at 30\,mJy and 50 target sources at 1\,mJy.
This results in a total of 10 pointings.


Each pointings is imaged separately using MFS and the bright source in the centre of each pointing
is deconvolved down to a threshold of 2\,mJy using CLEAN.
Each image is then primary beam corrected.
The final images have noise of 0.5\,mJy/beam in the centre.

\subsubsection{Bright source flux density distribution}
Data set type 10 is analogous to data set 5 for VLA and aims to complicate deconvolution with a more realistic 
distribution of bright sources.
The flux densities for bright sources are randomised using the model by \cite{bethermin2012} for the 1.1\,mm sky.
This model is based on predictions for galaxy populations,
and agrees well with recent ALMA observations \citep{hatsukade2013}.
The \cite{bethermin2012} model results in 282 sources over an area of $100\arcsec\times100\arcsec$.

Except for the bright sources, data set type 10 is based on data set type 8.
Generation of faint source and imaging is done in the same way.

\subsection{Sparse array}
Finally, we also perform a simulation using a sparser array.
The array is based on the VLA A configuration but removes all but 8 dishes.
We will refer to this configuration as \emph{sparse array}.

Using this spares array we produce data set type 11,
which is otherwise based on data set type 5.
To compensate for the lower number of baselines the integration time has been increased to a total of 600,000 seconds or approximately 7 days.
This means that the thermal noise limit is similar for data set type 5 and 11.
As such the flux density distribution of the target sources is set to the same as for data set 5, 
i.e., a Schechter distribution with $\alpha = -1.6$ and  $S_* = 1.7$\uJy{}.

%

\renewcommand{\arraystretch}{1.5}
\begin{table*}
	\centering
	\begin{minipage}{300mm}
		\caption{\label{tab:sim}Overview of the simulations and tests done}
		\begin{tabular}{@{}>{\centering\arraybackslash}m{12mm}cc>{\raggedright\arraybackslash}m{40mm}>{\raggedright\arraybackslash}m{32mm}p{4cm}@{}}
		\hline
		\hline
		Data set type & Telescope & Map size             &  Bright sources                                 & Target sources                                   & Main studied effect      \\

		\hline
		1 & VLA          & $  5\arcmin\times  5\arcmin$  & 1 point source with flux density of 5 mJy       & 25  point sources with flux densities of 6\,\uJy & Basic VLA observation   \\
		2 & VLA          & $  5\arcmin\times  5\arcmin$  & 1 point source with flux density of 5 mJy       & 25  point sources with flux densities of 6\,\uJy & Lower resolution imaging \\
		3 & VLA          & $ 30\arcmin\times 30\arcmin$  & 10 point sources with flux densities of 5 mJy   & 900 point sources with flux densities of 1\,\uJy & Wide-field effects       \\
		4 & VLA          & $ 30\arcmin\times 30\arcmin$  & 10 point sources with flux densities of 5 mJy   & 900 point sources with Schechter distribution, 
		                                                                                                    $\alpha = -1.6$, $S_* = 1.71$\uJy                 & Flux density distribution of target sources\\
		5 & VLA          & $ 30\arcmin\times 30\arcmin$  & Log polynomial distribution of flux densities, 
		                                                   417 sources                                     & 900 point sources with Schechter distribution, 
						   								                                                     $\alpha = -1.6$, $S_* = 1.71$\uJy                & Flux density distribution of bright sources\\
		6 & VLA          & $ 30\arcmin\times 30\arcmin$  & Log polynomial distribution of flux densities, 
		                                                   417 sources with 
		                                                   a Gaussian spatial size distribution 
		   				   							       with a full width half maximum varying 
		   				   							       from 0\fas5 to 5\arcsec.                        & 900 point sources with Schechter distribution,
														                                                     $\alpha = -1.6$, $S_* = 1.71$\uJy                & Extended bright sources \\ 
		7 & VLA          & $ 30\arcmin\times 30\arcmin$  & Log polynomial distribution of flux densities,
		                                                   417 sources                                     & 900 sources with a flux of 2.96\uJy{} 
														                                                     and a Gaussian spatial extension of 1\fas5.      & Extended target sources    \\
		8 & ALMA         & $100\arcsec\times100\arcsec$  & 10 point sources with flux densities of 30 mJy  & 50 point sources with 1.0 mJy flux density each  & Contiguous ALMA mosaic     \\
		9 & ALMA         & $  5\arcmin\times  5\arcmin$  & 10 point sources with flux densities of 30 mJy  & 50 point sources with 1.0 mJy flux density each  & Non-contiguous ALMA mosaic \\
		10& ALMA         & $100\arcsec\times100\arcsec$  & Sources generated with Bèthermin et al. 
		                                                   model (~282 sources with brightest 
														   flux density at ~10mJy)                         & 50 point sources with 0.2 mJy flux density each  & Flux density distribution of bright sources\\
		11& sparse array & $ 30\arcmin\times 30\arcmin$  & Log polynomial distribution of flux densities, 
		                                                   417 sources                                     & 900 point sources with Schechter distribution, 
														                                                     $\alpha = -1.6$, $S_* = .5\,$mJy                 & Sparse uv-coverage \\
		\hline
	\end{tabular}
	\end{minipage}
\end{table*}
\renewcommand{\arraystretch}{1.0}

\section{STACKING ALGORITHMS}
\label{sec:algorithms}

In the literature, several image domain based methods have been used for stacking of interferometric data.
\cite{carilli2008} stack sources in the VLA survey of the COSMOS field and adopt a median stacking method to this data,
where image pixels are calculated as the median of the same pixel in each sub image.
\cite{pannella2009} based their work on the same data set,
but improve on the method by fitting a dirty beam after median stacking.
\cite{decarli2014} chose a different approach for a non-contiguous ALMA mosaic. 
They adopted a weighted mean stacking,
where the weights are calculated as the square of the primary-beam attenuation.
All these methods share in common that they work on fully imaged data
and in this work we will refer to this group of stacking algorithms as \emph{image-stacking}.

Image-stacking represents an adaptation of stacking algorithms from non-interferometric telescopes.
It does not take into consideration the nature of interferometric data.
Since the $uv$-coverage of a real telescope does not cover the whole $uv$-plane,
it is not possible to image interferometric data without interpolating between sampled $uv$-points.
As such, imaging combined with a deconvolution algorithm provide us with a best guess for our actual sky,
but it may introduce artefacts that are not present in the $uv$-data.
To avoid such issues, it is in many cases preferable to work directly on the actual data: the visibilities.
In this work we refer to the stacking algorithms working directly on the visibilities as \emph{$uv$-stacking}.

In this paper, we present a novel $uv$-stacking method.
For comparison, we have also implemented image-stacking methods, both median based, 
similar to the algorithm used by \cite{pannella2009} and weighted mean based,
similar to the algorithms used by \cite{decarli2014}.

\subsection{General procedure and basic definitions}
\label{sec:genproc}
Stacking relies on the fact that we know the positions of our target sources
through observations of the same sources at a different wavelength.
In this work, we will not focus on the details of how the positions of the target sources are determined.
For each stacking algorithm, we assume the existence of a list of positions where we know targets to be present.
We refer to these positions as the \emph{stacking positions}.
When we stack at these positions, we arrive at a stacked source that exhibits the average properties of the target sources.

All targets are expected to have a flux density no greater than the noise in the data.
However, in a typical data set we would have other brighter sources present in the data.
To ensure that these sources do not greatly impact the result,
it is important to remove them before stacking.

As such, the overall procedure of stacking is:
\begin{enumerate}
	\item{} Produce a model of all sources that are visible in the data set 
	        and are not part of the target sources.
	        Subtract the model from the data to produce a residual data set.
	\item{} Perform stacking on the residual data set using the stacking positions.
	\item{} Determine properties of the stacked source.
\end{enumerate}


%

\subsection{Image-stacking}
\label{sec:imagestack}
For each stacking position,
we cut out a small square which we refer to as a \emph{stamp}.
The stamp is centred at the pixel closest to the stacking position 
and typically has a size of $64\times64$ pixels ($20\times20$ beams).
These stamps are stacked on a pixel by pixel basis using either median or weighted averaging.
For weighted averaging, the weights are calculated as
\begin{equation}
	W_k = 1/\sigma_k^2
	\label{eq:weights}
\end{equation}
where $\sigma_k$ is an estimate of the locale noise at stacking position $k$.

For VLA images, we calculate $\sigma_k$ from the image map.
For each stamp,
we mask out a circle in the centre with a radius equal to the major axis of the beam.
The noise is then calculated as the brightness standard deviation of the remainder of the stamp.
This noise estimate includes effects from primary beam as well as noise from residuals of bright sources.
It is important that the stamp is large enough to estimate the noise,
we use a size that covers approximately 300 beam areas.

In the case of ALMA maps,
using the same method as for VLA data to calculate $\sigma_k$ is not possible.
Due to the small primary beam of ALMA,
it is difficult to define a sufficiently large stamp to estimate the noise with the same method as for VLA.
However, the noise will typically not be dependant on dynamic range,
since the small field of view means that bright sources can be avoided in extra-galactic surveys.
As such, to estimate the local noise for ALMA,
we use $\sigma_k = 1/A_N(\bmath{\hat{S}_k})$, where 
$A_N(\bmath{\hat{S}})$ describes the primary-beam attenuation in the direction $\bmath{\hat{S}}$,
and $\bmath{\hat{S}_k}$ is the direction of stacking position $k$.
The primary-beam attenuation is calculated using the models included in CASA 4.2.

None of the target stacked sources are deconvolved. 
This results in a stacked image where the average of the sources is convolved with the dirty beam.
For point-sources we use the peak to estimate the flux density,
this is independent of the convolution.

\subsection{$uv$-stacking}
\label{sec:uvstack}
The idea of $uv$-stacking is to work directly on the visibilities.
\cite{hancock2011} used an \emph{ad hoc} approach to stack emission from supernovae in nearby galaxies in the $uv$-domain.
They achieve this by concatenating visibilities for each supernova.
This method is not possible to apply to large extra galactic surveys due to numerical limitations of current computers,
e.g. the \cite{carilli2008} stacking would generate a stacked data set of approximately 40TB.
Meaning stacking would be significantly more numerically challenging than cleaning the original data set.
For the much larger data sets of VLA and ALMA this would be even further out of reach.

We adopt a method where sources within the same pointing are not duplicated
but the visibilities recalculated using
\begin{equation}
	V_\mathrm{stack}(u,v,w) = V(u,v,w) \frac{\Sigma_{k=1}^N W_k \frac{1}{A_N(\bmath{\hat{S}_k})} e^{\frac{2\pi}{\lambda} i \bmath{B}\cdot\left( \bmath{\hat{S}_0} - \bmath{\hat{S}_k} \right)} 
	}{\Sigma_{k=1}^N W_k}
	\label{eq:uvstack}
\end{equation}
where $\bmath{\hat{S}_k}$ is a unit vector pointing to the stacking positions, 
$\bmath{\hat{S}_0}$ is a unit vector pointing to the phase centre, 
$\bmath{B}$ is the baseline of the visibility,
$\lambda$ is the wavelength, 
and $W_k$ is the weight of the stacking position.
The vector $\bmath{\hat{S}_k}$ is parametrised in the 
coordinates $(l_k, m_k)$, the coordinates relative to the phase centre in the east and north direction.
This changes $\bmath{B}\cdot\left( \bmath{\hat{S}_0} - \bmath{\hat{S}_k} \right)$ to $u l_k + v m_k + w(\sqrt{1-l_k^2-m_k^2}-1)$.

Stacking using this method does not increase the size of the data set.
Since the computation for each visibility is independent,
the code can be parallellized and run quickly for large data sets.
The weights are calculated using the same local noise estimators as in image-stacking.
Note that $A_N(\bmath{\hat{S}_k})$ enters both in $W_k$ and directly in equation \ref{eq:uvstack},
this ensures that the post-stacking flux density is corrected for primary beam attenuation.

\subsubsection{Flux density estimate}
From the stacked data set we want to estimate the flux density of the source at the phase centre, 
which is the stacked source.
For point sources we calculate this as the average of all non-flagged visibilities weighted by the visibility weights.
It is also possible to image the source, 
and we will do this for illustration,
but the imaged version is not used to estimate flux density.

\subsubsection{Mosaics}
The $uv$-stacking method as described only handles data sets with a single pointing.
For mosaics we run the stacking for each pointing separately according to equation \ref{eq:uvstack}.
Since all stacking sources are now shifted into the phase centre of each pointing,
the pointings can be concatenated into one pointing.
This results in a new data set with the same size as the combination of the pointings before stacking.

When concatenating visibilities it is important to ensure the 
visibility weights ($W^\mathrm{vis}_{j,k} (t)$ for visibility at the baseline between antenna $j$ and $k$ at time $t$)
are proper.
For our $uv$-data set,
the number of stacking positions may not be constant over all pointings.
When using $uv$-stacking,
visibility weights needs to be corrected,
to ensure that pointings with more stacking positions are given higher weights.
We achieve correct relative weights by applying

\begin{equation}
	W^\mathrm{vis, stack}_{j,k} (t) = \left( \sum_l W_l \right) W^\mathrm{vis}_{j,k} (t) 
\end{equation}

where $W_l$ are the weights of stacking positions in the pointing.
Note that for mosaics,
$W_l$ will not be the same in different pointings,
since they depend on $A_N(\bmath{\hat{S}_l})$ which in turn depends on the pointing centre.



\subsubsection{Wide-field effects}
Equation \ref{eq:uvstack} applies the exact phase (and delay) shifts to the visibilities,
accounting for the curvature of the celestial sphere.
Hence,
the phase corrections computed from equation \ref{eq:uvstack} can be applied to interferometric observations with any field of view.
However, 
a practical limitation in the equation is that the $(u,v,w)$ baseline coordinates,
for each source used in the stacking,
are not re-projected in accordance to the phase shifts.
As a consequence, for sources with large separations to the phase-centre of the observations,
the real $(u,v,w)$ coordinates of the baselines will differ from those computed at the phase-centre,
hence mapping into small shifts in Fourier space that are not taken into account in equation \ref{eq:uvstack}.

We note, though, that this limitation is only important when the
Fourier transform of the stacked sources strongly depends on the
$(u,v)$ coordinates. For the stacking of point-like sources (i.e.,
when the visibilities are independent of the $(u,v)$ coordinates), 
equation \ref{eq:uvstack} can be applied with no restrictions.

In the case of stacking of extended sources, we have estimated
that the baseline re-projection effects should not be larger than a
few per cent of the baselines lengths, at most.
For instance in the case of VLA observations at 1.4\,GHz in extended configuration 
(where these effects are larger),
a field of view of 30\arcmin{} at an elevation of 45$^\circ$ translates into changes of baseline length of 0.9 per cent.
These biasing effects in the baseline lengths translate directly into biases in the estimated average size of the stacked sources,
which are also of the order of a few per cent at most.

\subsection{Estimate of sizes for extended sources}
To estimate the size of extended sources,
we fit a model to our stacked source.
In the case of $uv$-stacking we use a standard $uv$-model fitting to a circular Gaussian,
i.e. only two degrees of freedom: flux density and size.
This is done by using {\tt uvmultifit} \citep{vidal2014}.

For model fitting in the image-domain we use the dirty beam.
We construct a fit function as a circular Gaussian convolved with the dirty beam plus a constant.
We will refer to this fitting of this function as PSF-fitting since it uses the point spread function (dirty beam).
This fit is done in the stacked image stamp.
We perform this fit by minimizing $\chi^2$ using the non-linear minimizer {\tt leastsq} of the {\tt scipy} package {\tt optimize}. 
\footnote{http://docs.scipy.org/doc/scipy/reference/\\/generated/scipy.optimize.leastsq.html}

\subsection{Evaluating noise }
When averaging over a large number of positions (sources), the noise will decrease.
Assuming that the different positions are statistically independent,
we can calculate this decreased noise as
\begin{equation}
	\sigma^2 = \frac{1}{\sum_{k=1}^N {1/\sigma_k^2}}
	\label{eq:noise_weighted}
\end{equation}
for the variance weighted mean, where $\sigma_k$ is the noise in position $k$.
If we approximate the noise at each position with a typical noise for the map, $\sigma_\mathrm{map}$, we arrive at the familiar
\begin{equation}
	\sigma = \frac{\sigma_\mathrm{map}}{\sqrt{N}},
	\label{eq:noise}
\end{equation}
where $N$ is the total number of stacking positions.

While we expect the stacking to roughly follow this trend we can not expect perfect agreement.
The primary beam attenuation and bright sources near to stacking positions will change the noise characteristics.
As such it is important to have a more reliable method to estimate noise in our stacked result.

\subsubsection{Simulated data sets}
\label{sssec:simnoise}
In this paper we study stacking using simulated data sets.
This allows to generate a large number of data sets.
For each data set type we generate 100 data sets.
For each data set the model is re-randomized,
ensuring that the data set have the same source distribution
but the noise is statistically independent.

Each data is stacked using both $uv$- and image-stacking.
This results in 100 stacked values for both $uv$- and image-stacking for each data set type.
From the distribution of these stacked values the noises of the stacking methods
are evaluated as the standard deviation.

\subsubsection{Real data sets}
\label{sssec:realnoise}
The method described in §\ref{sssec:simnoise} is not intended to be used for real data sets, 
but to provide the most reliable noise estimate possible to evaluate the stacking algorithms.
When working with real data sets,
we do not have the possibility to generate multiple data sets.
In this case method to estimate the noise using only one data set is needed.

For real data sets, noise can be estimated using a Monte Carlo method,
where we introduce fake sources into the data set with the same flux density and numbers to actual target sources.
The fake sources are added to the residual $uv$-data set directly using equation \ref{eq:uvmod}.
These sources are stacked and the process is repeated with several sets of fake sources.
The distribution of these stacked flux densities will tell us the distribution,
we can expect when stacking real sources.
We can estimate the noise of our real stacked flux density as the standard deviation of our fake flux densities.
Furthermore, we can use this method to estimate if there is an offset in the flux density;
we refer to this as bias in the following sections.
We apply this method to our simulated data sets of type 6 to test the accuracy of the noise and bias estimates.

\section{RESULTS}
\label{sec:results}

\begin{figure}
	\vbox to90mm{\vfil \includegraphics[width=85mm]{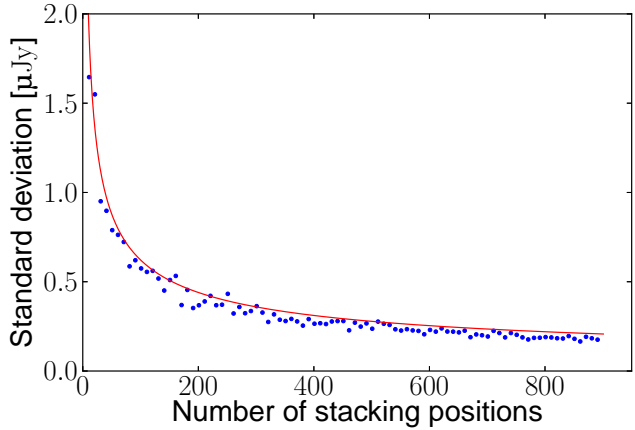}
	\caption{\label{fig:numvnoise}The noise of stacked flux density as a function of number of target sources using a data set of type 6. 
	Red line indicates the expected $1/\sqrt{N}$ falloff expected from stacking independent positions.
	}
	\vfil}
\end{figure}

The results from stacking of simulated data are summarised in table \ref{tab:simresults}.
Average is calculated over 100 data sets for each data set type 
and noise is calculated as the scatter over the 100 data sets (see \S \ref{sssec:simnoise}).
In general we find, as expected, that the noise is proportional to one over the square of the number of stacking, see Fig. \ref{fig:numvnoise}.

\begin{figure}
	\vbox to130mm{\vfil \includegraphics[width=85mm]{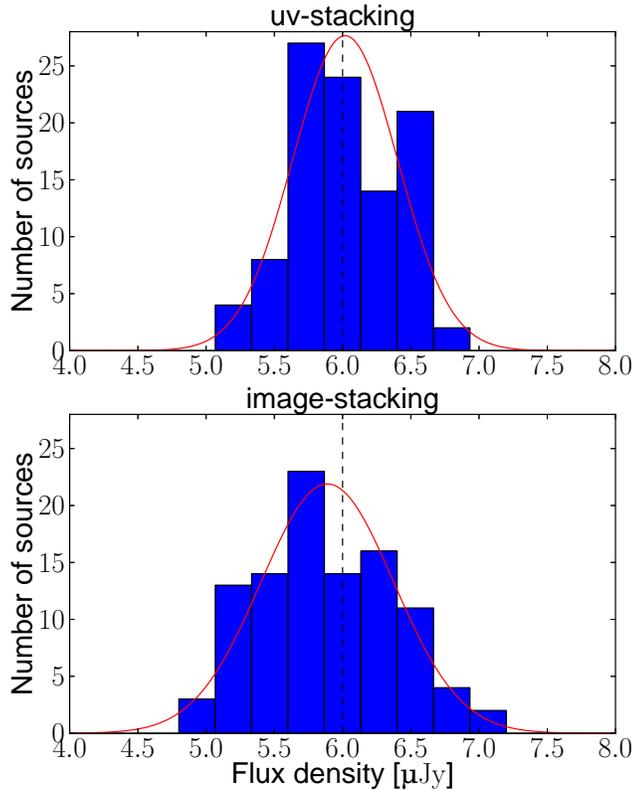}
	\caption{\label{fig:histset1}Distribution of estimates of mean flux density of the target population for data set type 1, 
		a simple VLA data set with small field of view and low dynamic range. 
		The image-stacking and $uv$-stacking produces equivalent Gaussian distributions.
		Dashed line indicates the true flux densities of the sources,
		in red the Gaussian distribution for the sample mean and standard deviation.}
	\vfil}
\end{figure}

\begin{figure}
	\vbox to140mm{\vfil \includegraphics[width=85mm]{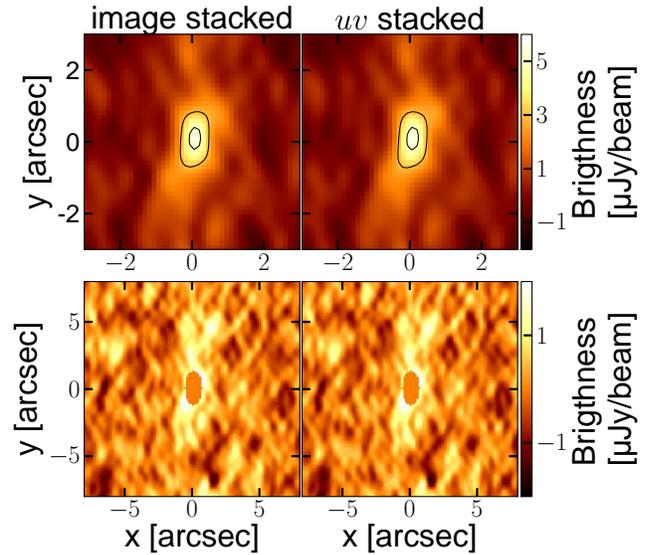}
	\caption{A typical stacked source from one simulation of data set type 1. (row 1) Stacked source. (row 2) Stacked stamp with central pixels masked. Contours are plotted at 3\uJy{} and 5\uJy.
	The central mask have a major axis of 5\fas89 and a minor axis of 2\fas94. 
	In none of the stamps the bright source has been deconvolved from the dirty beam.}
	\label{fig:set1_1_stackim}
	\vfil}
\end{figure}


\subsection{VLA}

\subsubsection{Basic case}
For data set type 1 image- and $uv$-stacking produce similar results. 
The distribution of flux density estimates from all 100 simulations can be seen in Fig. \ref{fig:histset1}.
The image-stacked flux density is typically 1 per cent lower than the $uv$-stacked flux density.
This is due to the pixel size of 0\fas25 that limits how well sources can be aligned in image-stacking.
The lack sub-pixel alignment of sources decreases the expected peak brightness for image-stacking to 5.92\uJy/beam,
which is the same as the measured 5.89\uJy/beam within the statistical uncertainty.

Fig. \ref{fig:set1_1_stackim} shows a typical image- and $uv$-stacked source.
As expected for data set type 1,
which has a very simple noise distribution,
the image- and uv-stacked stamps are very similiar.
Note that the stacked sources are not deconvolved from their dirty beams,
however, 
for uv-stacking the full uv data is available,
and could easily be deconvolved using a standard clean algorithm.


\subsubsection{Sub-pixel sampling}
For data set type 2 the pixel size is increased to 0\fas5.
This does not have any measurable impact on $uv$-stacking,
but for image-stacking it will further limit pixel alignment in stacking.
That results in the flux density of image-stacking is lowered by around 5 per cent down to 5.72\uJy{}.

\subsubsection{Wide field effect}
Data set type 3 spreads the sources evenly over a square area of $30\arcmin\times30\arcmin$.
This is covered within one pointing with VLA with a maximum beam attenuation of $0.25$ at the corners.
The $uv$-stacking yields a signal-to-noise ratio (SNR) that is approximately 10 per cent higher than that of image-stacking.

\subsubsection{Varying the flux density distribution of the target sources}
Data set type 4 uses a Schechter distribution ($\alpha = -1.6$ and $S_* = 1.71$\uJy) for the target population.
The mean of this distribution is 0.988\uJy.
The result for data set type 4 is very similar to data set type 3.
This indicate that the distribution of flux density of the target sources does not affect the stacking as long as the average is kept constant.

\subsubsection{Varying the flux density distribution of the bright sources}
\begin{figure}
	\vbox to120mm{\vfil \includegraphics[width=85mm]{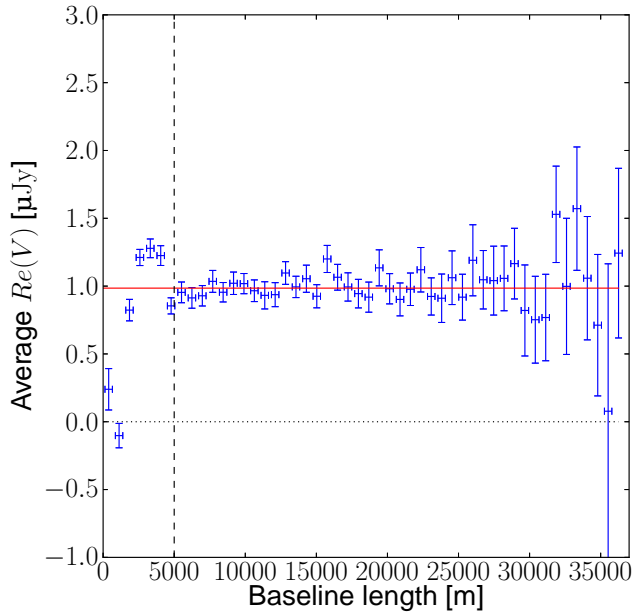}
	\caption{\label{fig:set5}
	         All $uv$-stacked visibilities of data set type 5 (combined from all simulations),
			 binned by baseline length.
			 For each bin, the flux density is calculated as the average of the real part of the visibilities.
			 The error bars are calculated as the standard deviation of the means 
			 of each simulation divided by the square root of the number of simulations.
			 The red curve show the average flux density of visibilities with baseline length over 5000m (dashed line).
			 For the shortest spacings,
			 the flux density is corrupted by incompletely subtracted bright sources.
	}
	\vfil}
\end{figure}

Data set type 5 contains a complex distribution of bright sources that affect the noise.
Using the Monte Carlo method described in \ref{sssec:realnoise} we find a negative bias for the stacked flux densities.
To look closer at this effect and study systematic effects in the $uv$-plane,
we bin the $uv$-stacked visibilities in bins of distance to the centre of the $uv$-plane ($\sqrt{u^2+v^2}$).
All simulations are averaged together, see Fig. \ref{fig:set5}.
We find that visibilities on short baselines, i.e. short uv distances deviate strongly from the expected values.
This effect is expected from imperfect removal of bright sources.
To avoid this issue,
we calculate the flux density in the $uv$-plane,
only using visibilities with spacings longer than 5000m (or 23.8 k$\lambda$).
Doing this,
we are able to reliably reproduce the flux density with $uv$-stacking.
The SNR with $uv$-stacking is 20 per cent higher than image-stacking

\subsubsection{Extended bright sources}
\begin{figure}
	\vbox to120mm{\vfil \includegraphics[width=85mm]{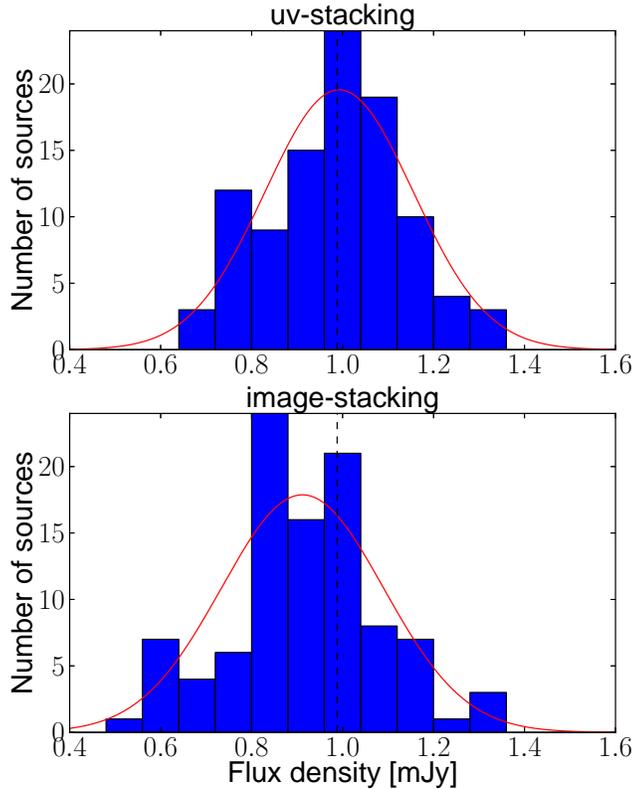}
	\caption{\label{fig:histset6}Distribution of estimated mean flux density for data set type 6,
	stacking of point sources in a map with extended bright sources.
	Dashed line indicates the true flux densities of the sources.
	In image-stacking the typical flux density is approximately 10\% lower than expected.
	}
	\vfil}
\end{figure}

In data set type 6 the noise is complicated further by the presence of extended bright sources.
We find a similar effect on the short baselines as was seen for data set type 5.
We exclude the shortest baselines (shorter than 5000m).
Not taking this into account during image-stacking leads to an expected bias in the flux densities measured using image-stacking. 
The flux densities measured with image-stacking are on average 10 per cent lower than the input model in our simulations, see Fig. \ref{fig:histset6}.


We use data set type 6 as a test case for estimating noise of a single data set.
This simulates how noise could be estimated for a real data set,
where we do not have access to 100 different data sets.
It is done by introducing additional fake sources into the data set (see \S \ref{sssec:realnoise}).
This results in a noise estimate of 0.16\uJy{} for $uv$-stacking and 0.18\uJy{} for image-stacking,
similar to the results found for the collection of data sets of type 5.

\begin{figure}
	\vbox to110mm{\vfil \includegraphics[width=85mm]{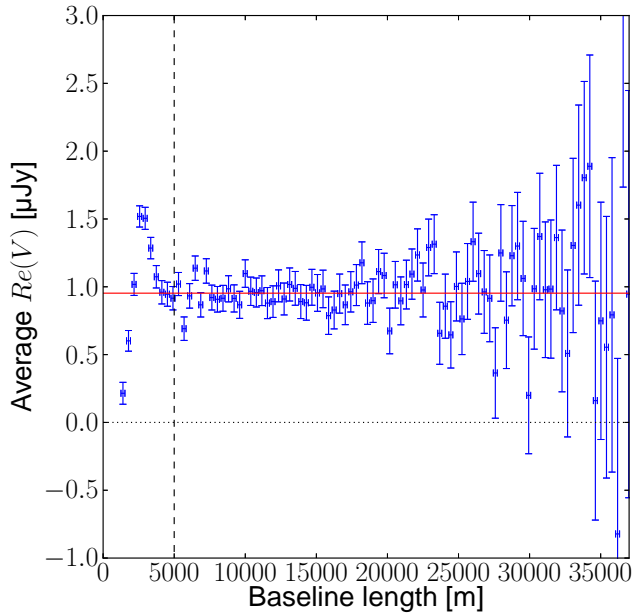}
	\caption{\label{fig:mc}
	Results from Monte Carlo noise and bias estimate.
	Random sources were introduced into one data set of type 6 and stacked.
	Average stacked visibilities have been binned by $\sqrt{u^2+v^2}$ and average real part of visibilities plotted.
	This clearly demonstrates the effects on short baselines by residuals of bright sources.
	Using this method it would be possible to find problematic data for real data sets.
	}
	\vfil}
\end{figure}

Using the random sources we can also study the effect of residuals on short baselines,
see Fig. \ref{fig:mc}.
This clearly shows the issues at below 5000m 
and demonstrates that having the stacked $uv$-data enable us to establish which baselines are robust,
and thereby derive a more reliable estimate.

\begin{figure}
	\vbox to110mm{\vfil \includegraphics[width=85mm]{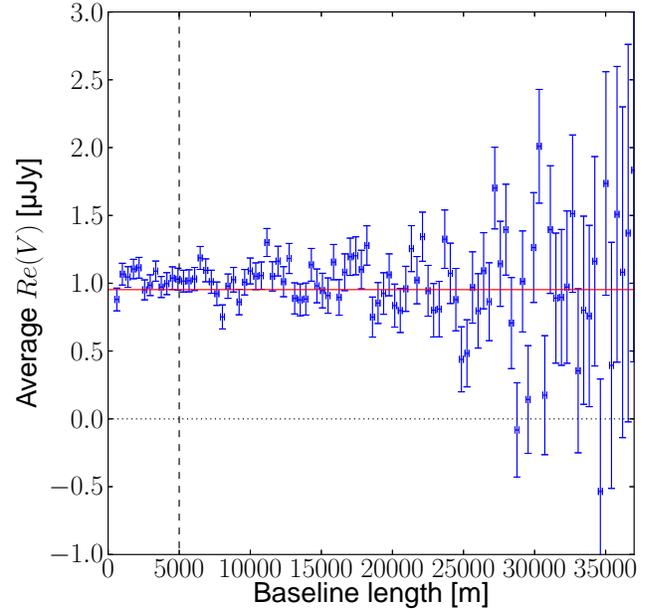}
	\caption{\label{fig:set6nob}
	         All $uv$-stacked visibilities of data set type 6 without bright sources,
			 binned by baseline length.
			 For each bin, the flux density is calculated as the average of the real part of the visibilities.
			 The error bars are calculated as the standard deviation of the means 
			 of each simulation divided by the square root of the number of simulations.
			 The red curve show the average flux density of visibilities (dashed line).
	}
	\vfil}
\end{figure}

To test if the issues on short baselines are related to incomplete removal of bright source,
a number of simulations of data set type 6 were run without introducing bright sources.
The thermal noise in these simulations was increased by a factor $\sqrt{2}$ to achieve a similar
noise level to the data sets with bright sources.
Fig. \ref{fig:set6nob} shows that no similar issues exist when no bright sources are present.


\subsubsection{Extended target sources}
\label{sec:res_ext_tar}
\begin{figure}
	\vbox to116mm{\vfil \includegraphics[width=85mm]{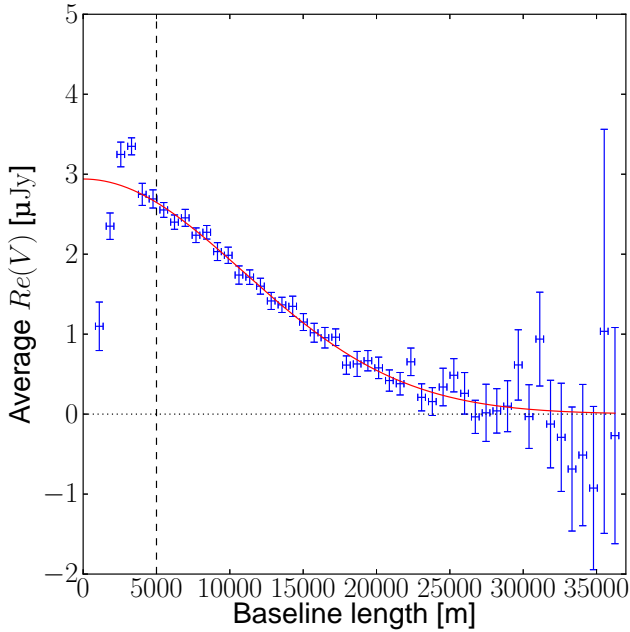}
	\caption{\label{fig:set7}
	         All $uv$-stacked visibilities of all simulations for data set type 7,
			 binned by $\sqrt{u^2+v^2}$. 
			 For each bin,
			 the flux density is calculated as the average of real 
			 part of the visibilities. The error bars are calculated as the 
			 standard deviation for each simulation divided
			 by the square root of the number of simulations. The red 
			 curve shows the average fitted size and flux densities,
		     using visibilities of baseline length longer than 5000m (dashed line).}
	\vfil}
\end{figure}

Data set type 7 introduces extended target sources. 
This requires that stacked flux densities are estimated through model fitting.
The increase in degrees of freedom in the result leads to an increase in noise on the measured flux density.
Fig. \ref{fig:set7} presents the average $uv$-stacking result for all simulations,
similar to Fig. \ref{fig:set5} for data set type 5.
This indicates a similar problem on short baselines as seen for data set 5.
As such we exclude all baselines shorter than 5000 m when estimating flux.

\begin{figure}
	\vbox to125mm{\vfil \includegraphics[width=85mm]{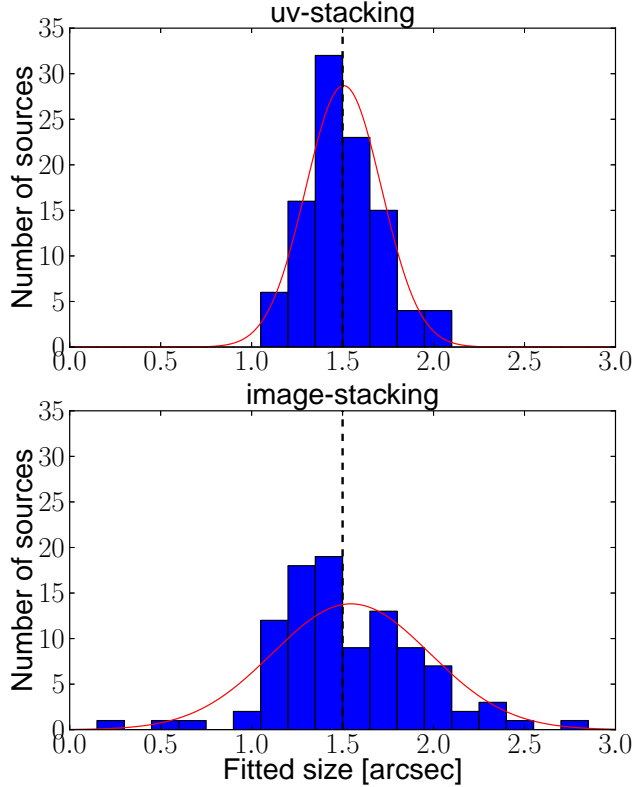}
	\caption{\label{fig:set7_sizes}
			 Distribution of fitted sizes for $uv$- and image-stacking for all realisations of data set type 7. 
			 Image-stacking estimates show a significantly broader distribution (standard deviation of $0\fas21$ versus $0\fas43$).}
	\vfil}
\end{figure}

For all data sets of this type we find an average flux density of 3.0$\pm$0.4\uJy{} and an average size of 1\fas50$\pm$0\fas2.
Where the noise is estimated as the standard deviation of our fitted parameters over 100 simulations.
With image stacking we find an average flux density of 3.0$\pm$0.5\uJy{} and an average size of 1\fas55$\pm$0\fas4.
In particular,
it can be seen that the size is significantly less accurate for image-stacking compared to $uv$-stacking,
with a noise more than double.
This appears to be more an effect on the shape on the source than the total flux density,
as the total flux density is only 30\% more noisy in image-stacking.
A comparison of the distribution of $uv$- and image-stacking size estimates can be seen in Fig. \ref{fig:set7_sizes}.

\begin{figure}
	\vbox to140mm{\vfil \includegraphics[width=85mm]{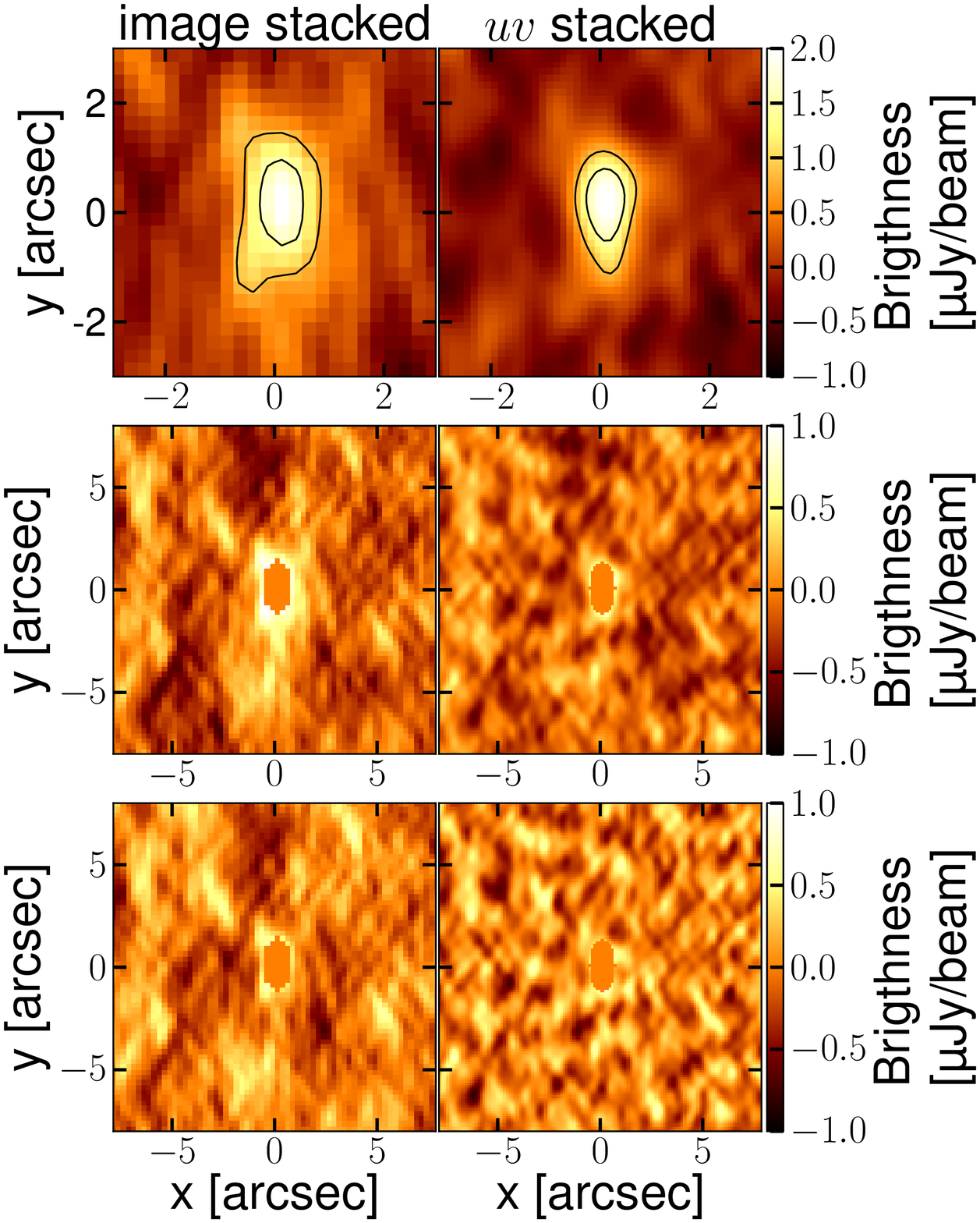}
	\caption{ \label{fig:set7_1_stamps}
	A typical stacked stamp from one simulatio of data set type 7. (row 1) Stacked source. Contours plotted at 0.9 and 1.5 \uJy{}.
	(row 2) Stacked stamp with central pixels masked out. 
	(row 3) Stacked stamp with central pixels masked out, and with all data at baselines shorter than 5 km masked.
	For image stacking the short baselines are masked by using fast fourier transform and setting central pixels to 0.
	The central mask is in all cases have a major axis of 5\fas89 and a minor axis of 2\fas94. 
	In none of the stamps is the bright source deconvolved.
	}
	\vfil}
\end{figure}

A typical stacked stamp is shown in Fig. \ref{fig:set7_1_stamps}.
The stamps can be seen to contain large scale variations resulting from residuals of nearby bright sources.
These variations are especially apparant in the image-stacked stamp.
For a source size of 1\fas5 convolved with the beam size in data set type 7,
the central mask blanks out the source down to a brightness of 40 per cent of peak, however,
for the image-stacked source, the brightness exceeds 40 per cent of the peak outside the mask.

In the third row of Fig. \ref{fig:set7_1_stamps} the short baselines are removed.
For uv-stacking this is implemented by ignoring baselines shorter than 5 km when imaging.
For image-stacking this is implemented by a fast Fourier transform (FFT) filter,
where the stamp is transformed with FFT,
the central pixels in the Fourier image are set to 0,
and Fourier image is transformed back with the inverse FFT.
Removing the short baselines results in a flatter noise distribution across stamps.
This is more effective for uv-stacked data,
while the large scale variations in the stamp are more severe for the filtered image-stacked stamp.
This is discussed further in section \ref{sec:dynrange}.

\subsection{ALMA}

\subsubsection{Contiguous mosaic}
For data set type 8,
the flux density is underestimated with image-stacking by 20 per cent.
This appears to be related to the use of mosaic mode in CASA.
We have found similar effects working with mosaic mode on other data sets.
Mosaic mode appears to blur sources towards the edges of the image,
and results in a stacked flux density systematically lower.
This indicates that image-stacking of CASA mosaics should be done with caution.

The noise of image-stacking is lower than $uv$-stacking for data set 8.
It should however be noted that since the image stacked signal is lower,
the SNR is the same for both.

\subsubsection{Non-contiguous mosaic}
Data sets of type 9 do not use mosaic cleaning,
since the pointings are not overlapping.
This yields a result that is similar for both uv and image-stacking.
There is still a small advantage for $uv$-stacking, compared to image-stacking, in regards to the noise.

\subsubsection{Varying the flux density distribution of the bright sources}
With data set type 10, 
we test a more complicated set of bright sources.
The effect here is not as pronounced as with data set type 5.
This is due to the target sources being closer in flux density to the bright sources.
In most ALMA surveys,
this will be true due to the small field of view of ALMA.
Result is that $uv$-stacking is more effective,
and yields a SNR 30 per cent higher than image-stacking.

\subsection{Sparse array}
With data set type 11, 
we test the effect of a significantly sparser $uv$-coverage.
In this case,
we find that the flux densities measured from image-stacking are significantly more scattered,
with a SNR 50 per cent higher than that of $uv$-stacking.

\begin{table*}
	\centering
	\begin{minipage}{140mm}
		\caption{\label{tab:simresults}Stacked flux densities for each set of simulations. Expected noise is calculated using equation \ref{eq:noise_weighted} with $\sigma$ calculated from the cleaned residual map. 
				 Flux density and noise are calculated as mean and standard deviation of the flux density from 100 simulations.
				 Stacking in the $uv$-plane can perform slightly better than the expected noise since it does not depend on the quality of the cleaned map.}

	\centering
	\begin{tabular}{c c c c }
		\hline
		\hline
		Data set type  & Expected value     & $uv$-stacking flux density (SNR)        & image-stacking flux density (SNR)   \\
		\hline
		1  & 6.00$\pm$0.47\uJy  & 6.02$\pm$0.38\uJy{} (15.8) & 5.89$\pm$0.49\uJy{} (12.0) \\
		2  & 6.00$\pm$0.47\uJy  & 6.02$\pm$0.38\uJy{} (15.8) & 5.73$\pm$0.49\uJy{} (11.7) \\
		3  & 0.99$\pm$0.12\uJy  & 0.97$\pm$0.12\uJy{}  (8.1) & 0.93$\pm$0.13\uJy{}  (7.2) \\
		4  & 0.99$\pm$0.13\uJy  & 0.98$\pm$0.12\uJy{}  (8.2) & 0.94$\pm$0.12\uJy{}  (7.8) \\
		5  & 0.99$\pm$0.12\uJy  & 0.99$\pm$0.13\uJy{}  (7.6) & 0.94$\pm$0.14\uJy{}  (6.7) \\
		6  & 0.99$\pm$0.15\uJy  & 0.99$\pm$0.16\uJy{}  (6.2) & 0.91$\pm$0.18\uJy{}  (5.1) \\
		7  & 2.96\uJy           & 2.96$\pm$0.41\uJy{}  (7.2) & 3.01$\pm$0.54\uJy{}  (5.6) \\
		8  & 1.00$\pm$0.10\,mJy & 0.98$\pm$0.12\,mJy   (8.2) & 0.90$\pm$0.11\,mJy   (8.2) \\
		9  & 1.00$\pm$0.09\,mJy & 1.00$\pm$0.08\,mJy  (12.5) & 0.97$\pm$0.09\,mJy  (10.8) \\
		10 & 0.20$\pm$0.02\,mJy & 0.20$\pm$0.02\,mJy  (10.0) & 0.17$\pm$0.03\,mJy   (5.7) \\
		11 & 0.99$\pm$0.32\uJy  & 1.00$\pm$0.20\uJy{}  (5.0) & 1.06$\pm$0.32\uJy{}  (3.3) \\
		\hline
	\end{tabular}
	\end{minipage}
\end{table*}

\section{DISCUSSION}
\label{sec:discussion}

Following the results of our simulations,
we find that the $uv$-stacking algorithm generally performs better,
and never worse,
than the image-stacking.
We discuss the main differences in the performance of the two algorithms.

In general, the results of image- and $uv$-stacking agree well.
In no case do they differ by more than 10 per cent in flux density
and the largest difference in SNR is 20 per cent.
However, the uv-stacking flux density is in all cases closer to the true average flux density of the simulated target sources compared to image-stacking.

Stacking in the $uv$-domain requires more time to run compared to image-stacking.
For the largest $uv$-data set (data set type 6) this required 20 minutes to run on a 
CPU-based server (12 cores).
While this is significantly more than the time required to do image-stacking,
it is also significantly less than the time required to image the same data set.

%

\subsection{Estimation of noise in stacked data}
An important consideration when interpreting stacked data is a good estimate of the noise in the data.
In this paper we have used a noise estimation algorithm which relied on generating multiple simulated data sets.
This is not possible for real data sets,
but other methods have also been presented in this algorithm.
The simplest method is to use the individual noise estimates for each position,
equation \ref{eq:noise_weighted}.
Noise estimate based on this is presented as expected noise in table \ref{tab:simresults}.
This noise estimate is typically accurate to 10 per cent.

A Monte Carlo noise estimate was evaluated for data set type 6,
where various random sources were introduced into the data set and noise estimated as the scatter
of the results of stacking these random sources.
This noise estimate produces a result that is the same as the simulation results within the statistical uncertainties.
However, this Monte Carlo noise estimate is computationally expensive.
Compared to the $uv$-stacking this method requires in the order of 50 times more time to run.
This can be improved by using a GPU based code in place of the CPU,
and we have evaluated this for a test case and found this to be approximately 150 times faster.

\subsection{Sub-pixel sampling}
Using image-stacking limits the alignment of sources to the pixel size used in imaging.
No such limitation exists for $uv$-stacking.
As such image-stacking will under-predict the flux densities and over-predict the sizes of the stacked sources.
Imaging with 3 pixels across the minor axis,
the flux densities measured in image-stacking will be 0.95 times the actual average flux densities of the target sources.
Our simulations confirm this to be the case, 
see especially the difference between data set type 1 and 2,
but the effect is present for all following VLA data sets (type 2-6).

\subsection{Wide-field effects}
The VLA field of view is around 30\arcmin at 1.4$\,$GHz.
Over such a large field of view, we must consider the 3D arrangement of the array.
If this is not taken into account sources will be incorrectly imaged towards the edges of the field.
In the case of image-stacking,
we worked with w-projection to produce our image map.
Our simulations show this to be effective for the VLA field of view.
With even larger field of views,
such as those of low frequency arrays such as LOFAR or the Murchison Widefield Array (MWA),
this may no longer be the case.
Using $uv$-stacking in these cases could be helpful,
since it simplifies the wide-field issues.

Equation \ref{eq:uvstack} is always correct for point-sources,
but because it does not have a transformation of the $(u,v)$-coordinates, 
there could be uncertainties for the extended sources.
As shown in Section \ref{sec:uvstack},
this is a small effect for VLA,
typically less than a few per cents.
For a larger field of view this error will be larger,
e.g. a 10$^\circ$ field of view observed at an elevation of 45$^\circ$,
the error would be $\sim{}20$ per cent.

If the size and shape of sources are important for a data set with large field of view,
it would be possible to modify our $uv$-stacking algorithm with faceting.
This would increase the size of the resulting data set,
but would allow us to handle much larger field of views.
Note again that this is not necessary for either ALMA or VLA,
even for large mosaics,
as the main determining factor is the field of view of each pointing.

\subsection{Bright foreground sources (dynamic range effects)}
\label{sec:dynrange}
For a typical deep VLA survey,
the sensitivity achieved will be such that the brightest sources will have a SNR of 1000 or more.
Sources of interest for stacking will have a SNR of a few or less.
As described in Section \ref{sec:genproc},
we subtract our model of the bright sources from the data.
However,
to be able to subtract the brightest sources,
we need very deep cleaning.
Doing so will generally introduce a ``clean bias'', 
which offsets the background level
(for a more detailed description see \citealt{condon98}).

We have used the clean models for our data (details in Section \ref{sec:genproc}),
although it would be possible to improve on the model by fitting the sources directly.
We have chosen not to do this,
since this would likely produce a better model
than would be possible for a real data set.

We see clear differences in how $uv$- and image-stacking handles the effect of residual bright sources.
For $uv$-stacking,
we have access to the full uv data after stacking.
Using this,
we have found that residuals mostly affect the shortest baselines.
By excluding these baselines,
we can ensure that $uv$-stacking suffers much less from dynamic range effects.
This becomes more pronounced as the distribution and structure of bright sources get more complicated.
When going from data set 5 to data set 6 with extended bright sources,
we find that the SNR decreases for both $uv$-stacking (from 7.6 to 6.2) and image-stacking (from 6.7 to 5.1).
It is clear that image-stacking suffers worse from this complication.
Our bright sources are Gaussian in the spatial extent.
For the complicated morphology of real world sources,
this effect may be even more pronounced.

\begin{figure}
	\vbox to80mm{\vfil \includegraphics[width=85mm]{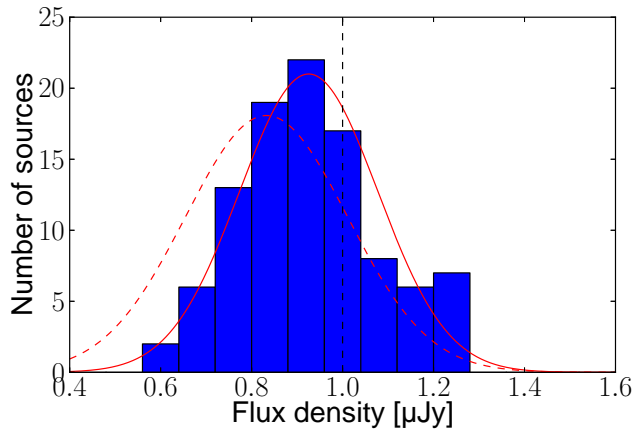}
	\caption{\label{fig:d_filt}Distribution of stacked flux densities for data set 6 using the $uv$-plane filter.
	Dashed Gaussian indicates distribution of non-filtered flux densities.
	Removal of extended components shifts the distribution closer to the true flux density,
	marked as the black dashed line.
	}
	\vfil}
\end{figure}

\begin{figure}
	\vbox to95mm{\vfil \includegraphics[width=85mm]{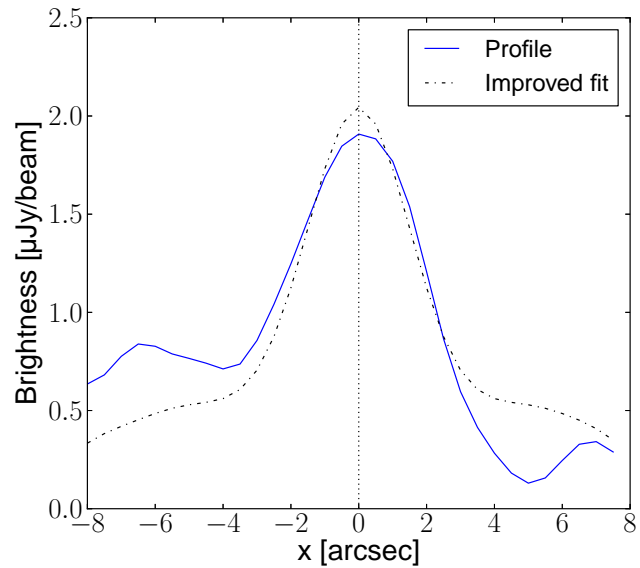}
	\caption{\label{fig:set7_profile}
		The blue line indicates a typical profile of stacked source from image stacking along the minor axis,
		i.e., a plot of the flux of pixels along the x-axis through the centre of the stamp.
		The dashed line indicates a fit to the full stamp with a Gaussian plus a constant bias level.
	}
	\vfil}
\end{figure}

When doing the $uv$-stacking,
it is possible to determine the problematic baselines.
As this is not possible for image-stacking in the same manner, 
we expect the larger bias in image-stacking to be related to this effect.
To test this,
we applied a low-spatial-frequency filter to remove these components from the image-stacked stamps in data set 6.
We stacked large stamps of 256x256 pixels (64\arcsec{}x64\arcsec{}) and Fourier transformed them.
We set the central 3x3 pixels in the Fourier image to zero and Fourier transformed back to the image-domain.
This removes baselines shorter than approximately 4500m.
The result of this was to change the flux density measurement from $0.91\uJy\pm0.18$ to $0.94\pm0.15$ (see Fig. \ref{fig:d_filt}),
a significant improvement, bringing the SNR of image-stacking up to only 10 per cent less than $uv$-stacking.
Fig. \ref{fig:set7_1_stamps} shows the effects of the filter when applied to data set type 7.
The difference seen between image- and uv-stacking can be understood as a combination of sub-pixel sampling and 
how short baselines are treated.

Even with filtering, we still see a 10 per cent difference.
This is probably due to two reasons.
First, the size of the stamp limits the pixel size in the $uv$-domain.
For a big stamp of 64\arcsec{}x64\arcsec{},
we are still limited to remove the few central pixels in the $uv$-plane.
This gives us less precision in setting which baselines are removed from the data.
Second, in image-stacking we are still limited by pixel size.
It is also important to note that the uv-stacking was important to determine the correct filtering scale.

\subsection{ALMA surveys}
ALMA observations and surveys will generally not be affected by wide-field effects,
due to the small antenna beam at ALMA frequencies.
The smaller field of view also means that the dynamic range will typically be lower,
since the probability of having bright sources in the same pointing as our target sources is smaller.
For our simulated ALMA data sets,
we have a typical flux density ratio of 30-50 between the brightest source and target sources,
compared to around 10,000 in data sets of type 7.

The differences between data sets of type 8 and 9 become apparent for image-stacking.
For data set type 9, each pointing is isolated and can be imaged and deconvolved separately.
Doing this results in image-stacking finding the right flux density with a SNR 8 per cent lower than $uv$-stacking.
For data set type 8, the mosaic is contiguous and is deconvolved with mosaic mode in CASA.
Using image-stacking produces a flux density that is 10 per cent too low 
compared to the true flux density,
but the SNR is very close for $uv$- and image-stacking.

Data sets of type 10 are also a contiguous mosaics,
but with a more complex distribution of bright sources,
and introducing a number of sources with intermediate flux densities.
This does not influence the result strongly,
but it still appears to be dominated by the same effect as in data set 8,
with image-stacking producing a flux density approximately 10 per cent lower.
In total, $uv$-stacking yields a SNR 30 per cent higher than image-stacking.

In some cases it may be interesting to stack sources in pointings with significantly higher dynamic range than those of data set type 8-10,
e.g., using archival data which has been targeted on a very bright source.
These cases have not been specifically been simulated in our analysis.
However, the effects of bright sources in ALMA data are expected to be similar to the effects of bright sources in VLA data.
In these cases $uv$-stacking is expected to present the same advantages as uv-stacking for data set type 6.
The small angular size of each pointing of ALMA will mean that the noise from bright sources will vary between different poitings.

To conclude with the discussion on ALMA,
we find that there are some advantages in using $uv$-stacking.
Especially in the case of contiguous mosaics,
where $uv$-stacking produces a much more reliable result.
However,
the difference is in general smaller than in the case of high dynamic range VLA data sets.

\subsection{Sparse interferometric array}
Both ALMA and VLA arrays consist of a relatively large number of antennas.
This results in a reasonably well covered $uv$-plane,
with no large sampling holes between the shortest and longest baseline.
However, 
for other telescopes, which also carry out large extragalactic surveys, e.g. e-MERLIN and ATCA, 
the number of antennas is much smaller (7 and 6),
and thus the $uv$-coverage much less complete.
For these sparser arrays,
it would be expected that the effect of large dynamic range on the stacking is more severe.
We have tested this by generating an artificial array from a subset of the VLA antennas.

Although the thermal noise limit is identical for data sets of type 5 and 11,
we find that the sparse array in data sets of type 11 results in a significantly higher noise in the deconvolved map
( from 0.11\uJy{} to 0.32\uJy{}.)
For $uv$-stacking in data set 11 we are excluding the shortest baselines.
Compared to data set type 5 the shortest baselines behave worse in data sets of type 11,
and as such,
the advantage of $uv$-stacking is even bigger.
The SNR is 50 per cent larger for $uv$-stacking compared to image-stacking.

%

\subsection{Modelling of stacked sources}
All the results in data set types 1-6 and 8-11 regard the study of stacking of point sources.
Realistically,
it is possible that some of the target sources are extended.
This will significantly complicate the estimate of flux denstiy.
We can no longer rely on peak brigtness as a flux density estimate.
We either need to fit the source to a model or integrate over a larger area.
In this work, we have estimated the flux density through model fitting.
This also allows us to estimate sizes of our stacked sources.

We found that the size estimates using $uv$-stacking were significantly more accurate that those derived from image-stacking,
with typical errors less than half for $uv$-stacking compared to image-stacking.
In flux density estimates the methods differ less.
The flux density error for image-stacking is 40 per cent higher than that for $uv$-stacking.

For $uv$-stacking,
we fit a Gaussian directly to the stacked visibilities.
However,
if the dynamic range is large,
the same effects are present as in data sets 5 and 6;
the shortest baselines must be removed from the data.

For image-stacking,
it is important to take some extra care when fitting.
There are two main complications present for image-stacking.  
Firstly, when fitting in the image domain our source will be convolved with a beam.
In the case of our algorithm we do not clean the source after stacking,
that is, the source is convolved with the dirty beam.
We use PSF-fitting to estimate the size of the actual source,
i.e., fitting the data to a model convolved with the dirty beam.

As a result of the convolution we are required to use a non-linear fitter for this problem.
When fitting in the $uv$-domain we can rewrite a Gaussian fit as a polynomial fit.
This can be solved with a linear fitter. 
This can be a large advantage,
both in performance for large data sets,
as well as in robustness for ill-conditioned fits.

The second complication is the residuals from bright sources in high dynamic 
range data sets.
As with point sources this effect is mitigated for $uv$-stacking by 
removing the shortest baselines.
For image-stacking this is not as easy.
As such there will often be an extended component in the image-stacked stamps, 
which could affect the model fitting.
We reduce this effect by also fitting a bias level in the image.
The need for this bias level can be seen in Fig. \ref{fig:set7_profile}.
Alternative it would be possible to use a low-spatial-frequency filter to remove this offset,
as described in Section \ref{sec:dynrange}.

Another important point is that model fitting in image-stacking
relies on knowing the dirty beam well.
For a mosaic data set observed at different times,
the $uv$-coverage could vary between pointings.
In this case it may be difficult to determine the dirty beam for
the final image, complicating model fitting in the image plane further.

\subsection{Stacking code}
The code used in this paper is available under GNU Public License through the Nordic ARC node (nordic-alma.se)
\footnote{Available under the name {\tt STACKER} as part of the ALMA nordic ARC software http://www.nordic-alma.se/support/software-tools}.
This code is in format of a CASA task and implements both the $uv$- and image-stacking algorithms of this paper.
On top of stacking code, it also provides several useful tasks for stacking related data processing.
It allows introducing and stacking of random sources to estimate bias and noise.
It implements the $uv$- and image-fitting algorithms used in the paper.
It also allows to remove a model of bright sources from the data.

\section{CONCLUSIONS}
Stacking of radio and mm interferometric data is a powerful tool to estimate the average properties of sources that are not individually detected.
We have carried out an extensive analysis of two algorithms for stacking of interferometric data:
$uv$-stacking which carries out the stacking directly on the visibilities in the $uv$-plane, 
and image-stacking which does this on the imaged data.
The latter has so far been the more common approach to date.

We find that $uv$- and image-stacking both are effective methods for stacking interferometric data
and that they produce similar results.
However, with $uv$-stacking we have access to the full $uv$-data post stacking.
This can be a large advantage compared to image-stacking.
For example,
for high dynamic-range data sets, 
this allows us to identify and reduce the effects of imperfect deconvolution of bright sources.
Simulation show this to be especially important for deep VLA surveys,
where the large dynamic range complicates the imaging.

For stacking extended sources,
we find that $uv$-stacking provides an even more significant advantage.
Full access to $uv$-data post stacking allows for model fitting.
This halves the uncertainty of stacked size estimates for VLA data.

Many of the issues, such as high dynamic range and wide field,
will become more dramatic with new telescopes.
For LOFAR and MWA,
the large field of view ensures that any pointing will contain sources close to the maximum possible brightness.
And for the Square Kilometre Array (SKA), the target depth needed for deep surveys is such that dynamic range will be a significant limitation.
For either of these telescopes,
having access to the $uv$-data after stacking will be invaluable in ensuring that the desired signal is optimally extracted from the data.

%
%
%

\section{Acknowledgements}
LL would like to thank Robert Beswick for fruitful discussion and helpful comments.
We thank an anonymous referee for constructive comments, 
which helped improve the manuscript.
KK acknowledges support from the Swedish Research Council.
\label{lastpage}

\bibliography{references} 

\begin{thebibliography}{29}
\expandafter\ifx\csname natexlab\endcsname\relax\def\natexlab#1{#1}\fi

\bibitem[{{Bartelmann} \& {White}(2003)}]{bartelmann03}
{Bartelmann} M., {White} S.~D.~M., 2003, \aap, 407, 845

\bibitem[{{B{\'e}thermin} {et~al}\mbox{.}(2012){B{\'e}thermin}, {Daddi},
  {Magdis}, {Sargent}, {Hezaveh}, {Elbaz}, {Le Borgne}, {Mullaney}, {Pannella},
  {Buat}, {Charmandaris}, {Lagache}, \& {Scott}}]{bethermin2012}
{B{\'e}thermin} M. {et~al.}, 2012, \apjl, 757, L23

\bibitem[{{Bondi} {et~al}\mbox{.}(2008){Bondi}, {Ciliegi}, {Schinnerer},
  {Smol{\v c}i{\'c}}, {Jahnke}, {Carilli}, \& {Zamorani}}]{bondi2008}
{Bondi} M., {Ciliegi} P., {Schinnerer} E., {Smol{\v c}i{\'c}} V., {Jahnke} K.,
  {Carilli} C., {Zamorani} G., 2008, \apj, 681, 1129

\bibitem[{{Brandt} {et~al}\mbox{.}(2001){Brandt}, {Hornschemeier}, {Alexander},
  {Garmire}, {Schneider}, {Broos}, {Townsley}, {Bautz}, {Feigelson}, \&
  {Griffiths}}]{brandt01}
{Brandt} W.~N. {et~al.}, 2001, \aj, 122, 1

\bibitem[{{Cady} \& {Bates}(1980)}]{cady1980}
{Cady} F.~M., {Bates} R.~H.~T., 1980, Optics Letters, 5, 438

\bibitem[{{Carilli} {et~al}\mbox{.}(2008){Carilli}, {Lee}, {Capak},
  {Schinnerer}, {Lee}, {McCraken}, {Yun}, {Scoville}, {Smol{\v c}i{\'c}},
  {Giavalisco}, {Datta}, {Taniguchi}, \& {Urry}}]{carilli2008}
{Carilli} C.~L. {et~al.}, 2008, \apj, 689, 883

\bibitem[{{Condon}(1992)}]{condon}
{Condon} J.~J., 1992, \araa, 30, 575

\bibitem[{{Condon} {et~al}\mbox{.}(1998){Condon}, {Cotton}, {Greisen}, {Yin},
  {Perley}, {Taylor}, \& {Broderick}}]{condon98}
{Condon} J.~J., {Cotton} W.~D., {Greisen} E.~W., {Yin} Q.~F., {Perley} R.~A.,
  {Taylor} G.~B., {Broderick} J.~J., 1998, \aj, 115, 1693

\bibitem[{{Conway}, {Cornwell} \& {Wilkinson}(1990){Conway}, {Cornwell}, \&
  {Wilkinson}}]{conway90}
{Conway} J.~E., {Cornwell} T.~J., {Wilkinson} P.~N., 1990, \mnras, 246, 490

\bibitem[{{Cornwell}, {Golap} \& {Bhatnagar}(2008){Cornwell}, {Golap}, \&
  {Bhatnagar}}]{cornwell2008}
{Cornwell} T.~J., {Golap} K., {Bhatnagar} S., 2008, IEEE Journal of Selected
  Topics in Signal Processing, 2, 647

\bibitem[{{Decarli} {et~al}\mbox{.}(2014){Decarli}, {Smail}, {Walter},
  {Swinbank}, {Chapman}, {Coppin}, {Cox}, {Dannerbauer}, {Greve}, {Hodge},
  {Ivison}, {Karim}, {Knudsen}, {Lindroos}, {Rix}, {Schinnerer}, {Simpson},
  {van der Werf}, \& {Wei{\ss}}}]{decarli2014}
{Decarli} R. {et~al.}, 2014, \apj, 780, 115

\bibitem[{{Dole} {et~al}\mbox{.}(2006){Dole}, {Lagache}, {Puget}, {Caputi},
  {Fern{\'a}ndez-Conde}, {Le Floc'h}, {Papovich}, {P{\'e}rez-Gonz{\'a}lez},
  {Rieke}, \& {Blaylock}}]{dole06}
{Dole} H. {et~al.}, 2006, \aap, 451, 417

\bibitem[{{Greaves} {et~al}\mbox{.}(2012){Greaves}, {Hales}, {Mason}, \&
  {Matthews}}]{greaves2012}
{Greaves} J.~S., {Hales} A.~S., {Mason} B.~S., {Matthews} B.~C., 2012, \mnras,
  423, L70

\bibitem[{{Hancock}, {Gaensler} \& {Murphy}(2011){Hancock}, {Gaensler}, \&
  {Murphy}}]{hancock2011}
{Hancock} P.~J., {Gaensler} B.~M., {Murphy} T., 2011, \apjl, 735, L35

\bibitem[{{Hatsukade} {et~al}\mbox{.}(2013){Hatsukade}, {Ohta}, {Seko}, {Yabe},
  \& {Akiyama}}]{hatsukade2013}
{Hatsukade} B., {Ohta} K., {Seko} A., {Yabe} K., {Akiyama} M., 2013, \apjl,
  769, L27

\bibitem[{{Hodge} {et~al}\mbox{.}(2013){Hodge}, {Karim}, {Smail}, {Swinbank},
  {Walter}, {Biggs}, {Ivison}, {Weiss}, {Alexander}, {Bertoldi}, {Brandt},
  {Chapman}, {Coppin}, {Cox}, {Danielson}, {Dannerbauer}, {De Breuck},
  {Decarli}, {Edge}, {Greve}, {Knudsen}, {Menten}, {Rix}, {Schinnerer},
  {Simpson}, {Wardlow}, \& {van der Werf}}]{hodge2013}
{Hodge} J.~A. {et~al.}, 2013, \apj, 768, 91

\bibitem[{{H{\"o}gbom}(1974)}]{Hogbom1974}
{H{\"o}gbom} J.~A., 1974, \aaps, 15, 417

\bibitem[{{Knudsen} {et~al}\mbox{.}(2005){Knudsen}, {van der Werf}, {Franx},
  {F{\"o}rster Schreiber}, {van Dokkum}, {Illingworth}, {Labb{\'e}},
  {Moorwood}, {Rix}, \& {Rudnick}}]{knudsen05}
{Knudsen} K.~K. {et~al.}, 2005, \apjl, 632, L9

\bibitem[{{Lehmer} {et~al}\mbox{.}(2007){Lehmer}, {Brandt}, {Alexander},
  {Bell}, {McIntosh}, {Bauer}, {Hasinger}, {Mainieri}, {Miyaji}, {Schneider},
  \& {Steffen}}]{lehmer2007}
{Lehmer} B.~D. {et~al.}, 2007, \apj, 657, 681

\bibitem[{{Mart{\'{\i}}-Vidal} {et~al}\mbox{.}(2014){Mart{\'{\i}}-Vidal},
  {Vlemmings}, {Muller}, \& {Casey}}]{vidal2014}
{Mart{\'{\i}}-Vidal} I., {Vlemmings} W.~H.~T., {Muller} S., {Casey} S., 2014,
  \aap, 563, A136

\bibitem[{{Miller} {et~al}\mbox{.}(2013){Miller}, {Bonzini}, {Fomalont},
  {Kellermann}, {Mainieri}, {Padovani}, {Rosati}, {Tozzi}, \&
  {Vattakunnel}}]{miller2013}
{Miller} N.~A. {et~al.}, 2013, \apjs, 205, 13

\bibitem[{{Miller} {et~al}\mbox{.}(2008){Miller}, {Fomalont}, {Kellermann},
  {Mainieri}, {Norman}, {Padovani}, {Rosati}, \& {Tozzi}}]{miller2008}
{Miller} N.~A., {Fomalont} E.~B., {Kellermann} K.~I., {Mainieri} V., {Norman}
  C., {Padovani} P., {Rosati} P., {Tozzi} P., 2008, \apjs, 179, 114

\bibitem[{{Pannella} {et~al}\mbox{.}(2009){Pannella}, {Carilli}, {Daddi},
  {McCracken}, {Owen}, {Renzini}, {Strazzullo}, {Civano}, {Koekemoer},
  {Schinnerer}, {Scoville}, {Smol{\v c}i{\'c}}, {Taniguchi}, {Aussel}, {Kneib},
  {Ilbert}, {Mellier}, {Salvato}, {Thompson}, \& {Willott}}]{pannella2009}
{Pannella} M. {et~al.}, 2009, \apjl, 698, L116

\bibitem[{{Planck Collaboration} {et~al}\mbox{.}(2014){Planck Collaboration},
  {Ade}, {Aghanim}, {Armitage-Caplan}, {Arnaud}, {Ashdown}, {Atrio-Barandela},
  {Aumont}, {Baccigalupi}, {Banday}, \& et~al.}]{planck2013}
{Planck Collaboration} {et~al.}, 2014, \aap, 571, A16

\bibitem[{{Schinnerer} {et~al}\mbox{.}(2007){Schinnerer}, {Smol{\v c}i{\'c}},
  {Carilli}, {Bondi}, {Ciliegi}, {Jahnke}, {Scoville}, {Aussel}, {Bertoldi},
  {Blain}, {Impey}, {Koekemoer}, {Le Fevre}, \& {Urry}}]{schinnerer2007}
{Schinnerer} E. {et~al.}, 2007, \apjs, 172, 46

\bibitem[{{Scoville} {et~al}\mbox{.}(2007){Scoville}, {Aussel}, {Brusa},
  {Capak}, {Carollo}, {Elvis}, {Giavalisco}, {Guzzo}, {Hasinger}, {Impey},
  {Kneib}, {LeFevre}, {Lilly}, {Mobasher}, {Renzini}, {Rich}, {Sanders},
  {Schinnerer}, {Schminovich}, {Shopbell}, {Taniguchi}, \&
  {Tyson}}]{scoville2007}
{Scoville} N. {et~al.}, 2007, \apjs, 172, 1

\bibitem[{{Steidel} {et~al}\mbox{.}(1999){Steidel}, {Adelberger}, {Giavalisco},
  {Dickinson}, \& {Pettini}}]{steidel1999}
{Steidel} C.~C., {Adelberger} K.~L., {Giavalisco} M., {Dickinson} M., {Pettini}
  M., 1999, \apj, 519, 1

\bibitem[{Thompson, Moran \& Swenson(2001)Thompson, Moran, \&
  Swenson}]{thompson}
Thompson A.~R., Moran J.~M., Swenson G.~W., 2001, Interferometry and Synthesis
  in Radio Astronomy; 2nd ed. Wiley-VCH, Weinheim

\bibitem[{{Webb} {et~al}\mbox{.}(2003){Webb}, {Eales}, {Foucaud}, {Lilly},
  {McCracken}, {Adelberger}, {Steidel}, {Shapley}, {Clements}, {Dunne}, {Le
  F{\`e}vre}, {Brodwin}, \& {Gear}}]{webb03}
{Webb} T.~M. {et~al.}, 2003, \apj, 582, 6

\end{thebibliography}
\end{document}